%% file: main.tex
\documentclass[sigconf]{acmart}

\input{preamble}

\AtBeginDocument{%
  }

\begin{document}

\title{Adequately Tailoring Age Verification Regulations}

\author{Shuang Liu}
\affiliation{%
  \institution{Carnegie Mellon University}
  \city{Pittsburgh}
  \country{United States}}
\email{shuangl4@andrew.cmu.edu}

\author{Sarah Scheffler}
\affiliation{%
  \institution{Carnegie Mellon University}
  \city{Pittsburgh}
  \country{United States}}
\email{sscheffl@andrew.cmu.edu}

\renewcommand{\shortauthors}{Shuang Liu and Sarah Scheffler}

\input{src/00_Abstract}


\maketitle

\input{src/10_Introduction}

\input{src/20_Current_Age_Verification_Legislation}

\input{src/30_Policy_analysis_model}

\input{src/40_Technical_Infra}

\input{src/50_Case_study}

\input{src/60_Challenges_Future_Direction}

\begin{acks}
We thank our CS\&Law shepherd and the anonymous reviewers for their thoughtful feedback and constructive suggestions, which substantially improved this work. This project was supported with funding from Carnegie Mellon University CyLab and from the Secure Blockchain Initiative. We are also grateful to all colleagues who provided informal feedback and discussion during the development of this research.
\end{acks}

\bibliographystyle{ACM-Reference-Format}
\bibliography{main}

\end{document}

%% file: preamble.tex
\usepackage{longtable}
\usepackage{makecell}
\usepackage{fontawesome}
\usepackage{algorithm}
\usepackage{algpseudocode}
\usepackage{hyperref}
\usepackage{xspace} 
\usepackage[normalem]{ulem} 
\usepackage{cleveref}
\usepackage{appendix}
\usepackage{xurl}

\usepackage{amssymb}
\usepackage{verbatimbox}
\usepackage{scalerel}
\usepackage{enumitem,kantlipsum}

\usepackage{hanging}
\usepackage{amsthm}

\usepackage{float}
\usepackage{multirow}
\useunder{\uline}{\ul}{}

\usepackage[table]{xcolor}   
\usepackage{booktabs}        
\usepackage{tabularx}        
\usepackage{array}          
\usepackage{caption}         
\usepackage{graphicx}        
\usepackage{colortbl} 
\newcolumntype{C}{>{\centering\arraybackslash}X}

\newcommand{\hlink}[1]{[\href{#1}{link}]}

\theoremstyle{plain}

\theoremstyle{definition}

\setcopyright{none}
\renewcommand\footnotetextcopyrightpermission[1]{}
\acmConference{}{}{}
\acmBooktitle{}
\acmYear{}
\acmISBN{}
\acmDOI{}

%% file: src/00_Abstract.tex
\begin{abstract}

In response to growing concerns about the harms to children online, especially regarding access to online pornography, a new wave of regulatory action has swept across the United States. To date, 25 states have implemented laws aimed at restricting minors’ access to such content by requiring websites to implement ``reasonable age verification.'' In June 2025, the Supreme Court’s decision in \textit{Free Speech Coalition v. Paxton} upheld the constitutionality of Texas H.B. 1181 -- one of the most constitutionally vulnerable of these age verification laws -- and found that it was subject to (and passed the standard of) intermediate scrutiny, with a need to ``adequately tailor'' age verification regulations.  
Yet, \textit{Free Speech Coalition v. Paxton} leaves unresolved practical challenges posed by state-level age-verification regimes. 

What is the state of age verification legislation in the US? How can the constitutional concept of ``adequate tailoring'' be interpreted in a way that is accessible and meaningful for non-legal experts, particularly those in technical and engineering domains who are attempting to interpret these laws in the context of technical trade-offs? What age-verification approaches are widely used today, what technical infrastructures and standards do they rely on, and how do they contribute to these concerns? This paper addresses those questions by proposing an analytical model to interpret ``adequate tailoring'' from multiple perspectives with associated governmental goals and interests, and properties of age verification systems, and by applying that model to evaluate both current state laws and widely used verification methods.  

This paper's major contributions include: (1) we map the current U.S. age-verification legislative landscape; (2) we introduce an analytical model to analyze ``adequate tailoring'' for age verification and potential application to other online regulatory policies; and (3) we analyze the main technical approaches to age verification, highlighting the practical challenges and tradeoffs from a technical perspective. Further, while we focus on U.S. state laws, the principles underlying our framework are applicable to age-verification debates and methods worldwide.

\end{abstract}

%% file: src/10_Introduction.tex
\section{Introduction}

In June 2022, Louisiana became the first U.S. state to require pornography websites to verify users’ ages beyond self-attested age \cite{Louisiana}. The law, effective January 1, 2023, sparked what some have called ``the New Pornography Wars'' \cite{dahlstrom2023new}. As of January 2026, 25 states have active age-verification laws \cite{Freespeech}. This wave of legislation reflects concerns over the impact of minors' ease of access to online pornography in the mobile internet era: All the laws specifically address ``sexual content appealing to the prurient interest''; some \textit{also} affect other forms of content perceived as harmful to minors, including nudity and depictions of sexual excitement or sadomasochistic abuse \cite{Indiana, Kansas}. 

Some parties, especially the Free Speech Coalition (FSC),  challenged these laws as unconstitutional restrictions on protected speech \cite{FSClitigation}. The most prominent case, \textit{Free Speech Coalition, Inc. v. Paxton} (henceforth \textit{FSC v. Paxton}) contested Texas H.B. 1181. On June 27, 2025, the Supreme Court upheld the law 6–3, applying \textit{intermediate scrutiny} and finding that the age verification under H.B. 1181 imposes an \textit{incidental} burden on adults’ protected speech while serving the state’s important interest in shielding children from harmful online content \cite{FreeSpeechCoalition_v_Paxton}. In particular, the Court emphasized that to pass intermediate scrutiny, ``a regulation is adequately tailored so long as the government's interest `would be achieved less effectively absent the regulation' and the regulation `does not burden \textit{substantially more speech than is necessary} to further that interest''' \cite[quoting \textit{TikTok, 604 U.S.}]{FreeSpeechCoalition_v_Paxton} (emphasis added).

The Supreme Court's ruling signals that mandatory age verification is a lasting feature of the online regulatory landscape in the U.S. However, constitutional survival does not resolve the practical and policy challenges these laws introduce, either in the U.S. or elsewhere in the world. Current statutes in U.S. states lean heavily on ID-based verification, with 16 states explicitly requiring age verification through ``submission of government-issued identification'', 
compelling disclosure of sensitive personal information that heightens risks of surveillance, data breaches, and chilling effects on lawful adult expression. Such tensions highlight the tradeoffs among child protection, privacy and security, customer convenience, vendor burdens, and other values.

In short, while \textit{FSC v. Paxton} established the application of ``intermediate scrutiny'' standards, in practice, several troublesome problems remain: How should ``adequate tailoring'' be interpreted, given different conflicting governmental interests? How can current state-level age verification laws be evaluated based on their performance on different legal objectives? What are the most suitable and tailored verification implementation methods? 

This paper addresses these issues by proposing an analytic model to interpret ``adequate tailoring'' with measurable variables and integrating technical analysis of ID-based verification methods with legal evaluation. More specifically, it focuses on the following research questions:
\begin{itemize}
    \item \textbf{RQ1:} What is the current state of affairs regarding age verification legislation in the US, and what context led here? To answer this question, we describe the current legislative landscape in \Cref{tab:comprehensivelegislation}, and situate it in the historical context of prior age-verification efforts in the U.S. (\Cref{sec:background})
    \item \textbf{RQ2:} How should ``adequate tailoring'' be interpreted? To address this question, we identify key governmental goals and tensions among four properties of age assurance systems: Assurance, Business Convenience, Consumer Convenience, and Data Protection together with related governmental interests. (\Cref{model})
    \item \textbf{RQ3:} What age-verification approaches are widely used today, what technical infrastructures and standards do they rely on, and how should these approaches be evaluated under the proposed model? In \Cref{technical} we summarize three categories of age verification approaches (direct ID submission, facial age estimation, and verifiable digital credentials) and analyze their tradeoffs across the four layers of the model introduced in \Cref{model}. (\Cref{technical})
    \item \textbf{RQ4:} What tradeoffs do current state-level age-verification laws present under this model? To answer this question, \Cref{legalanalysis} evaluates state statutes by identifying common issues and through a case study of the Tennessee Protecting Children from Social Media Act. (\Cref{legalanalysis})
\end{itemize}
The major contributions of this paper are as follows: 
First, it provides the \emph{first} comprehensive evaluation of both state-level age-verification laws and technical age verification approaches. 
Second, we propose a model for interpreting constitutional scrutiny regarding ``adequate tailoring'', including legitimate governmental interests and contradictory properties of the system. This model has potential applicability beyond age verification to other regulatory contexts (e.g. digital IDs). 
Third, we apply this model to analyze common expected trade-offs, and in a focused case study to examine how to evaluate one particular age verification law in the context of adequate tailoring.

The purpose of this paper is not to advocate for or against age verification, any specific statute, or any particular technology. Rather, it examines how policymakers should practically ``adequately tailor'' regulations and laws involving speech, privacy, and identity. 
Age verification is, to put it mildly, a heavily politicized topic in the U.S. presently. These laws have been described as pro-censorship, anti-LGBTQ, anti-feminist, anti-pornography, a moral panic, and as a vehicle for surveillance. Opponents to the laws have been described as promoting harm to children, porn apologists, privacy absolutists, or protecting the profits of tech companies. The political salience of age verification ensures it will be contested and repurposed across multiple agendas, with ongoing battles over the scope of its reach.
Age verification and \emph{FSC v. Paxton} represent major changes in how the U.S. regulates online speech, and its effects will be felt widely.
We also suspect more upcoming changes are likely on the way to verify other attributes in a similar way to age verification, which will also  likely be considered under the ``intermediate scrutiny'' standard.
Thus it is of critical importance to understand what this standard means in practice and how to apply it going forward.

%% file: src/20_Current_Age_Verification_Legislation.tex
\section{Background and Context to U.S. Age Verification Legislation}
\label{sec:background}

In this section we describe the current state of U.S. State-level age verification legislation, and provide a brief summary of important age verification context leading up to this point.

\input{table/comprehensive_legislation_table}

\subsection{Current U.S. State-Level Age Verification Legislation Landscape}

Recent state legislation reflects a rapidly consolidating approach to age verification, with a clear emphasis on enforcing minimum age thresholds—most commonly 18 years—through structured verification mechanisms. Table \ref{tab:comprehensivelegislation} summarizes state statutes that specify age verification requirements and related privacy provisions within those statutes.
States nominally permit a variety of verification approaches, and most states (eighteen) permit use of any ``reasonable commercial method'' to verify age. However, many states instead (or additionally) mention specific methods.
In practice, excluding the general ``reasonable commercial method'' language, the regulatory landscape is dominated by ID-based methods.
Of the twenty-five states with active age verification laws,
sixteen states, including Arizona and South Dakota accept submission of traditional IDs \cite{Arizona, Southdakota}, while thirteen states like Arkansas, Georgia, Indiana accept digital IDs \cite{Arkansas, Georgia, Indiana}. 
By contrast, specific mentions of non-ID-based approaches within the statute itself are far less common. Only three states, Nebraska, South Dakota, and Wyoming accept financial documents, while 18 states like Virginia accept commercial systems based on transactional data or other databases \cite{Virginia}, and nine states like Florida accept other methods \cite{Florida}.

Almost all of the statutes also address privacy and data retention (or non-retention) requirements. All states except Virginia \cite{Virginia} have at least some form of privacy requirement.  Eighteen states such as Alabama, Arkansas, and Florida, prohibit the retention of personal information after an age verification check has been completed \cite{Alabama, Arkansas, Florida}. Five states including Arizona, Indiana, Missouri, South Dakota, and Texas prohibit retention of verification data at any time \cite{Arizona, Arkansas, Missouri, Southdakota, Texas}.

\subsection{Protecting Children from Harmful Internet Content: From \emph{Reno} to \emph{Paxton}}
Although proof of age has been required for activities such as voting and marriage \cite{cross2004age} for many years, the legislative emphasis on protecting children (from harmful content or from other harms) arose much later \cite{marsden2023age}.
Modern debates on age verification in the U.S. trace back to the Communications Decency Act of 1996 and the landmark Supreme Court decision in Reno v. American Civil Liberties Union \cite{marsden2023age, Reno}. The First Amendment of the US Constitution protects individuals from government restrictions on expression, and courts establish three levels of scrutiny (i.e., ``strict scrutiny'', ``intermediate scrutiny'' and ``rational basis review'') when reviewing laws that limit protected speech. In Reno, the Court struck down provisions restricting knowing ``transmission'' of ``indecent'' or ``patently offensive'' online speech to minors, holding that they imposed unconstitutional burdens on free expression \textit{for adults} and failed in passing ``strict scrutiny'' \cite{Reno}. 
Since then, the tension between protecting children and safeguarding First Amendment rights has remained a contentious issue in both legislative and judicial arenas \cite{Thierer2004}. 
Apthorpe, Frischmann, and Shvartzshnaider examine the state of age gating before \emph{FSC v. Paxton} \cite{frischmann2024online}.

Despite this long history, the current wave of anti-pornography legislation -- and their court approval in \textit{FSC v. Paxton} -- departs from earlier efforts in several key respects \cite{FreeSpeechCoalition_v_Paxton}. First, the ubiquity of mobile internet access and the rise of advanced AI technologies have heightened public fears of online harms to minors. Violent or non-consensual pornography is widely available, and AI-generated sexual content, including deepfakes, can be accessed by minors with ease \cite{NYTDFnews, DFnews2, DFnews3}. Second, unlike Reno, where the Communications Decency Act of 1996 directly regulated speech content (e.g. banning transmission of ``indecent'' or ``patently offensive'' online speech) and thus was held overbroad and unconstitutional \cite{Reno}, today’s state laws focus on access conditions —requiring platforms to implement verification mechanisms to avoid direct content-based regulation \cite{anduze2024obscenity}.  
The third difference between current efforts and Reno is that the Communications Decency Act had no exception for material that had literary, artistic, educational, or other value for minors. On the other hand, among the new state age verification laws, most do have this exception. 
Finally, in Reno, a sticking point was that it was technically infeasible to perform ubiquitous age verification in a way that would still allow adults to view content. The perception of legislators and the courts has changed.

%% file: table/comprehensive_legislation_table.tex
\begin{table*}[ht]
  \centering
  \caption{State-Level Age Verification Laws: Methods and Privacy Requirements}
  \label{tab:comprehensivelegislation}
  \small
  \setlength{\tabcolsep}{2.5pt}
  \resizebox{\linewidth}{!}{%
  \begin{tabular}{@{}l l l c cccccc cccc@{}}
    \toprule
    & & & & \multicolumn{6}{c}{\textbf{Accepted Age Verification Methods*}}  & \multicolumn{4}{c}{\textbf{Privacy Requirements**}} \\
    \cmidrule(lr){5-10} \cmidrule(lr){11-14}
    \textbf{State} & \textbf{Bill} & \textbf{Effective Date} & \rotatebox{90}{\makecell{\textbf{Age}}} & \rotatebox{90}{\makecell{\textbf{Physical/}\\\textbf{Traditional ID}}} & \rotatebox{90}{\makecell{\textbf{Digital}\\\textbf{ID}}} & \rotatebox{90}{\makecell{\textbf{Photo}\\\textbf{Match}}} & \rotatebox{90}{\makecell{\textbf{Financial}\\\textbf{Document}}} & \rotatebox{90}{\makecell{\textbf{Commercial}\\\textbf{System***}}} & \rotatebox{90}{\makecell{\textbf{Other}\\\textbf{Methods****}}} & \rotatebox{90}{\makecell{\textbf{No Retention}\\\textbf{After Check}}} & \rotatebox{90}{\makecell{\textbf{No Retention}\\\textbf{Anytime}}} & \rotatebox{90}{\makecell{\textbf{Anonymous}\\\textbf{Method}}} & \rotatebox{90}{\makecell{\textbf{No Related}\\\textbf{Requirements}}}\\
    \midrule
    Alabama \cite{Alabama} & HB 164  & Oct. 1, 2024 & 18 & \faSquareO & \faSquareO & \faSquareO & \faSquareO & \faSquareO & \faCheckSquareO & \faCheckSquareO & \faSquareO & \faSquareO & \faSquareO\\
    Arizona \cite{Arizona} & HB 2112 & Sep. 25, 2025 & 18 & \faCheckSquareO & \faSquareO & \faSquareO & \faSquareO & \faCheckSquareO & \faSquareO & \faSquareO & \faCheckSquareO & \faSquareO & \faSquareO\\
    Arkansas \cite{Arkansas} & SB 66 & Jul. 1, 2023 & 18 & \faCheckSquareO & \faCheckSquareO & \faSquareO & \faSquareO & \faSquareO & \faCheckSquareO & \faCheckSquareO & \faSquareO & \faSquareO & \faSquareO\\
    Florida \cite{Florida} & HB 3 & Jan. 1, 2025 & 18 & \faSquareO & \faSquareO & \faSquareO & \faSquareO & \faSquareO & \faCheckSquareO & \faSquareO & \faSquareO & \faCheckSquareO & \faSquareO\\
    Georgia \cite{Georgia} & SB 351 & Jul. 1, 2025 & 16 & \faCheckSquareO & \faCheckSquareO & \faSquareO & \faSquareO & \faSquareO & \faCheckSquareO & \faCheckSquareO & \faSquareO & \faSquareO & \faSquareO\\
    Idaho \cite{Idaho} & H 498 & Jul. 1, 2024 & 18 & \faCheckSquareO & \faCheckSquareO & \faSquareO & \faSquareO & \faCheckSquareO & \faSquareO & \faCheckSquareO & \faSquareO & \faSquareO & \faSquareO\\
    Indiana \cite{Indiana} & SB 17 & Aug. 16, 2024 & 18 & \faCheckSquareO & \faSquareO & \faSquareO & \faSquareO & \faCheckSquareO & \faSquareO & \faSquareO & \faCheckSquareO & \faSquareO & \faSquareO\\
    Kansas \cite{Kansas} & SB 394 & Jul. 1, 2024 & 18 & \faSquareO & \faSquareO & \faSquareO & \faSquareO & \faSquareO & \faCheckSquareO & \faCheckSquareO & \faSquareO & \faSquareO & \faSquareO\\
    Kentucky \cite{Kentucky} & HB 278 & Jul. 15, 2024 & 18 & \faCheckSquareO & \faSquareO & \faSquareO & \faSquareO & \faCheckSquareO & \faSquareO & \faCheckSquareO & \faSquareO & \faSquareO & \faSquareO\\
    Louisiana \cite{Louisiana} & HB 142 & Jan. 1, 2023 & 18 & \faCheckSquareO & \faCheckSquareO & \faSquareO & \faSquareO & \faCheckSquareO & \faSquareO & \faCheckSquareO & \faSquareO & \faSquareO & \faSquareO\\
    Mississippi \cite{Mississippi} & SB 2346 & Jul. 1, 2023 & 18 & \faCheckSquareO & \faCheckSquareO & \faSquareO & \faSquareO & \faCheckSquareO & \faSquareO & \faCheckSquareO & \faSquareO & \faSquareO & \faSquareO\\
    Missouri \cite{Missouri} & 15 CSR 60-18 & Nov. 30, 2025 & 18 & \faCheckSquareO & \faCheckSquareO & \faSquareO & \faSquareO & \faCheckSquareO & \faSquareO & \faSquareO & \faCheckSquareO & \faSquareO & \faSquareO\\
    Montana \cite{Montana} & SB 544 & Jan. 1, 2024 & 18 & \faCheckSquareO & \faCheckSquareO & \faSquareO & \faSquareO & \faCheckSquareO & \faSquareO & \faCheckSquareO & \faSquareO & \faSquareO & \faSquareO\\
    Nebraska \cite{Nebraska} & LB 1092 & Jul. 1, 2024 & 18 & \faCheckSquareO & \faCheckSquareO & \faSquareO & \faCheckSquareO & \faCheckSquareO & \faSquareO & \faCheckSquareO & \faSquareO & \faSquareO & \faSquareO\\
    North Carolina \cite{NorthCarolina} & HB 8 & Jan. 1, 2024 & 18 & \faSquareO & \faSquareO & \faSquareO & \faSquareO & \faCheckSquareO & \faCheckSquareO & \faCheckSquareO & \faSquareO & \faSquareO & \faSquareO\\
    North Dakota \cite{Northdakota} & HB 1561 & Aug. 1, 2025 & 18 & \faCheckSquareO & \faCheckSquareO & \faSquareO & \faSquareO & \faCheckSquareO & \faSquareO & \faCheckSquareO & \faSquareO & \faSquareO & \faSquareO\\
    Ohio \cite{Ohio} & HB 96 & Sep. 30, 2025 & 18 & \faCheckSquareO & \faSquareO & \faSquareO & \faSquareO & \faCheckSquareO & \faSquareO & \faCheckSquareO & \faSquareO & \faSquareO & \faSquareO\\
    Oklahoma \cite{Oklahoma} & SB 1959 & Nov. 1, 2024 & 18 & \faSquareO & \faCheckSquareO & \faSquareO & \faSquareO & \faCheckSquareO & \faSquareO & \faCheckSquareO & \faSquareO & \faSquareO & \faSquareO\\
    South Carolina \cite{Southcarolina} & HB 3424 & Jan. 1, 2025 & 18 & \faSquareO & \faCheckSquareO & \faSquareO & \faSquareO & \faCheckSquareO & \faSquareO & \faCheckSquareO & \faSquareO & \faSquareO & \faSquareO\\
    South Dakota \cite{Southdakota} & HB 1053 & Jul. 1, 2025 & 18 & \faSquareO & \faSquareO & \faSquareO & \faCheckSquareO & \faSquareO & \faCheckSquareO & \faSquareO & \faCheckSquareO & \faSquareO & \faSquareO\\
    Tennessee \cite{Tennessee} & SB 1792 & Jan. 1, 2025 & 18 & \faSquareO & \faSquareO & \faCheckSquareO & \faSquareO & \faCheckSquareO & \faSquareO & \faCheckSquareO & \faSquareO & \faSquareO & \faSquareO\\
    Texas \cite{Texas} & HB 1181 & Sep. 1, 2023 & 18 & \faCheckSquareO & \faCheckSquareO & \faSquareO & \faSquareO & \faCheckSquareO & \faSquareO & \faSquareO & \faCheckSquareO & \faSquareO & \faSquareO\\
    Utah \cite{Utah} & SB 287 & May 3, 2023 & 18 & \faSquareO & \faCheckSquareO & \faSquareO & \faSquareO & \faCheckSquareO & \faSquareO & \faCheckSquareO & \faSquareO & \faSquareO & \faSquareO\\
    Virginia \cite{Virginia} & SB 1515 & Jul. 1, 2023 & 18 & \faSquareO & \faSquareO & \faSquareO & \faSquareO & \faCheckSquareO & \faCheckSquareO & \faSquareO & \faSquareO & \faSquareO & \faCheckSquareO\\
    Wyoming \cite{Wyoming} & HB 43 & Jul. 1, 2025 & 18 & \faSquareO & \faSquareO & \faSquareO & \faCheckSquareO & \faSquareO & \faCheckSquareO & \faCheckSquareO & \faSquareO & \faSquareO & \faSquareO\\
    \bottomrule
  \end{tabular}
  }
  \vspace{3pt}
  \begin{minipage}{\linewidth}
    \small
    \textit{Note:} \\
    1. *For Accepted Age Verification Methods, \faCheckSquareO~= methods explicitly accepted as an option; \faSquareO~= methods not explicitly accepted.\\
    2. **For Privacy Requirements, \faCheckSquareO~= strictly required in law; \faSquareO~= not required in law\\
    2. ***Commercial systems include those based on IDs, transactional data or other database, and other sources. \\
    3. ****Other methods accepted by law.
  \end{minipage}
\end{table*}

%% file: src/30_Policy_analysis_model.tex
\section{A Framework for Interpreting ``Adequate Tailoring''}
\label{model}

\textit{FSC v. Paxton} established the application of ``intermediate scrutiny'' standards for age verification: ``a regulation is adequately tailored so long as the government's interest would be achieved less effectively absent the regulation, and the regulation does not burden \emph{substantially more speech than necessary} to further that interest'' \cite[quoting \textit{TikTok, 604 U.S.}]{FreeSpeechCoalition_v_Paxton} (emphasis added).

However, how should websites, age verifiers, judges, and the public interpret  ``adequate tailoring'', especially given all those different governmental interests? In this section, we propose the analysis model to provide a structured way to reason about ``adequate tailoring''. 
We first identify the government's \textbf{Goals/Interests} by reading the statutes and surrounding contextual information (\Cref{sec:governmentalinterest}).
Then, we identify properties of age verification systems that interact with those goals. In \Cref{sec:system-properties} we identify four properties -- Assurance, Data Protection, Business Convenience, and Consumer Convenience -- which we will later use to understand what is being optimized, what is being sacrificed, and how alternative approaches compare. 

We emphasize that we are neither contesting the ruling in \textit{FSC v. Paxton} nor seeking to substitute constitutional scrutiny with purely quantitative analysis. Rather, our approach follows the Supreme Court’s decision while aiming to present a systematic interpretation of this complex, specialized form of scrutiny in a way that is accessible to non-constitutional law experts, including verification implementers and the broader public. In addition, the four-layers and governmental objectives proposed in this framework have broader potential applications in three respects. First, the four-layers remain useful even if the applicable level of scrutiny shifts -- for example, to strict scrutiny or to rational-basis review. Second, the framework can extend beyond age-verification laws to other types of access restrictions (say, verifying that someone is not a bot, or that someone is a resident in a particular area). Third, it not only supports legal analysis of constitutionality of laws but can also inform the selection of age verification \textit{implementations} by evaluating trade-offs among different implementation approaches in light of the relevant layers and goals.

\subsection{Categories of Government Goals}
We identified the following categories of governmental interests in age verification laws:
\label{sec:governmentalinterest}
\begin{itemize}
    \item \textbf{Prevent Children's Access to Harmful or Pornographic Material Goal} is the most commonly stated goal of age verification legislation is to limit minors’ access to pornography.
    For example, Alabama states that its law is intended ``to prevent pornography exposure and addiction to minors'' \cite{Alabama}. Similarly, Mississippi describes its objective as ``to regulate pornographic media exposure to children'' \cite{Mississippi}, while Arkansas seeks to establish liability for the ``publication or distribution of material harmful to minors on the internet'' \cite{Arkansas}
    \item \textbf{Child Protection Goal}. Some statutes articulate a more generalized child-protection interest without detailing specific categories of content. Kentucky, for instance, states its purpose simply as ``relating to the protection of children'' \cite{Kentucky}, and Wyoming similarly frames its law as ``relating to child protection'' \cite{Wyoming}. These formulations emphasize safeguarding minors’ welfare broadly, rather than targeting pornography alone. 
    \item \textbf{Privacy and Security Goal} denotes the governmental interest in preserving individual privacy and safeguarding data security while pursuing other objectives. For instance, most age verification laws impose requirements to avoid retaining personal information after an age check or to prohibit ongoing data storage (see \Cref{tab:comprehensivelegislation} ``Privacy Requirements'').
    \item \textbf{Consumer Rights Protection Goal} denotes the protection of lawful consumer rights. For instance, both Oklahoma and Kansas mention ``consumer protection'' in the age verification bills \cite{Oklahoma, Kansas}. Generally, in the age-verification context, Consumer Rights Protection Goal includes safeguarding the rights of adults to access online pornography and other legally permissible materials.
\end{itemize}
We suspect some policymakers also have other important governmental objectives not captured by the three categories above. For instance, lawmakers may also seek to preserve well-functioning digital marketplaces, avoid chilling lawful speech, and allow space for technical innovation in age-verification solutions. However, these interests were not explicitly stated in statutory text. 

\subsection{System Properties}
\label{sec:system-properties}
We identify four system properties of age verification systems that support one or more of the goals from \Cref{sec:governmentalinterest} -- but that also sometimes trade off against each other, capturing tensions among different legislative goals. 
\begin{itemize}
    \item \textbf{Assurance} denotes the probability that a mechanism correctly identifies a target at the point of access. In the context of age verification, Assurance corresponds to the success rate of correctly detecting minors who attempt to access online pornography or other harmful content, with particular concern for instances in which minors are erroneously classified as adults and granted access. 
    
    \item \textbf{Business Convenience} denotes the ease with which online platforms and other businesses can implement or procure a mechanism that achieves Assurance. For age verification, Business Convenience reflects the additional costs and burdens associated with implementing such systems (including loss of legitimate users who choose to avoid the site due to the age verification).
    \item \textbf{Consumer Convenience} denotes the ease with which consumers can access an online service while maintaining Assurance. In the age-verification setting, Consumer Convenience decreases as lawful adults experience greater friction or exclusion from access (willingly or otherwise).
    \item \textbf{Data Protection} denotes the degree to which information privacy and data security are preserved in achieving Assurance. In age verification, Data Protection diminishes as more sensitive information (such as identity attributes, biometrics, or device/network metadata) is collected or inferred, and as the potential for linkability increases.
\end{itemize}

\subsection{Statutory Constraints on System Properties}
\label{section3.3}
The four system properties, Assurance, Business Convenience, Consumer Convenience, and Data Protection—are not independently optimizable. State age-verification statutes impose recurring legal constraints that condition how these properties can be implemented and evaluated. These constraints are threshold statutory features, not downstream implementation choices, and they recur across jurisdictions. 

\subsubsection{Statutory ``reasonableness'', Assurance, and Business Convenience}
Most statutes permit compliance through ``reasonable'' or ``commercially reasonable'' age-verification methods without specifying accuracy thresholds or error tolerances  (e.g., \cite{Arizona, Georgia, Indiana, Kentucky}). This language functions as statutory interpretation rather than technical guidance, leaving Assurance legally indeterminate. 
``Reasonableness'' language is common in law and can in different settings refer to non-excessiveness, justifiability, non-incorrectness, relevance to purpose, practicality, or other meanings \cite{zipursky2015reasonableness}.
While the reasonableness language provides a useful way to avoid overly restrictive or potentially-easily-outdated rules, the absence of explicit requirements also lowers Business Convenience.
Implementers must design systems without knowing what level of under- or over-inclusiveness will later be deemed sufficient, shifting legal risk onto platforms and vendors. 

\subsubsection{Residency and location as constraints on Business Convenience}

Many statutes apply only to minors who are residents of, domiciled in, or physically present within a state for a specified duration. These predicates determine whether age verification is required at all, imposing an ex ante compliance burden on platforms. For example, Kentucky’s age verification framework applies to those sojourning for over 31 consecutive days \cite{Kentucky}, which maps poorly onto online access models, and common technical proxies such as IP-based geolocation do not reliably establish them. Because platforms must resolve these jurisdictional questions before triggering verification, residency requirements primarily burden Business Convenience and condition downstream trade-offs among the other properties.

\subsubsection{Data Protection conflicts with state privacy law}
Age-verification mandates frequently coexist with state privacy statutes that restrict the collection and retention of sensitive data, including government identifiers, biometrics, and precise location. This creates a structural tension: enforcing age-gated access often requires processing the very data privacy laws seek to minimize. Even where non-retention or anonymization is required, initial collection may still conflict with data-minimization and necessity principles.

%% file: src/40_Technical_Infra.tex
\section{Age Verification Methods}
\label{technical}

This section categorizes current prominent strong age verification approaches (i.e., direct submission of ID copies, verifiable digital credentials, and facial age estimation). Specifically, this section provides a review of the technical infrastructure and workflow, technical standards, industrial implementation, and privacy/security features of each verification method, and evaluates the systemic issue of identity-binding vs. device-binding and compares the privacy risks associated with on-device vs. remote age checking.

\subsection{Technical Infrastructures and Features}

We summarize recently-popularized age verification systems in this section, and we identify the properties of a \emph{typical} offering in such a system.

\subsubsection{Direct ID document upload}
\label{directID}
One of the most straightforward age verification methods is the direct upload of an official ID document. In this approach, users are prompted to provide a scan or photograph of a government-issued identification (e.g. driver’s license, passport) via a website or app. The system’s infrastructure typically includes an image capture interface and a document verification service, which will technically extract data from the ID using optical character recognition (OCR) or barcode scanning~\cite{idscan_ocr}. For example, U.S. driver’s licenses encode personal data in a PDF417 barcode per AAMVA standards, which can be read to obtain the date of birth and other details~\cite{AAMVAstandards}.

\paragraph{Analysis} Despite its effectiveness at directly checking an official birth date, the direct ID upload method comes with notable privacy and security concerns. On the privacy side, uploading a government ID exposes personal data well beyond just age (name, address, ID number, etc.), raising the stakes for data handling. Regarding assurance, service providers typically authenticate an uploaded ID copy through a pipeline that includes an automated check, human reviews, or both. At submission, systems run image-quality and tamper checks (e.g., detection of blurring, irregularities in fonts/spacing/edges, and evidence of digital editing), followed by template-based validation against known ID layouts and security features for the claimed issuing jurisdiction (e.g., expected placement of fields, UV/hologram regions—where visible—or microprint proxies)~\cite{detectfakeID}. Many vendors then parse the machine-readable zone: for U.S. driver’s licenses this commonly means decoding the AAMVA PDF417 barcode to cross-check the barcode payload (date of birth, name, ID number, address) against the human-readable fields and to flag internal inconsistencies that are common in forgeries~\cite{AAMVAstandards}. Where available, higher-assurance implementations also seek issuer-backed validation such as cryptographic signatures previously signed by the ID issuance authorities. In practice, however, uploaded physical license images often cannot be cryptographically verified end-to-end: many states’ barcodes do not contain the optional DMV-signed cryptographic digital signature. As a result, human review is frequently used as a backstop for borderline cases (e.g., low-confidence authenticity scores or suspected synthetic images), but manual inspection is inherently variable and can be defeated by high-quality counterfeits or well-crafted digital manipulations.

\subsubsection{Verifiable digital credentials}
Verifiable digital credentials are digital attestations and statements about an individual and their identity, that can be presented to a relying party and verified without relying on the relying party's direct access to an underlying identity database~\cite{verifiabledigitalcredential}. 
From the verification side, the verification process involves cryptographic verification.
However, on the credential presentation side, there are two main technical archetypes: Trusted Attestation of Selective Disclosure (Trusted Notaries), and Cryptographic Zero-Knowledge Proofs (ZKPs).

\paragraph{Trusted Attestation of Selective Disclosure}
In this method, the presentation is a \emph{signature by a party that is trusted by both the user and the verifier}. In both these cases, we refer to the trusted party as a \emph{Notary}, as the party's purpose is to provide assurance to some claim.
Possibilities for that trusted notary include:
\begin{itemize}
\item a Digital ID application provided by the government issuing the ID (e.g. the myColorado app \cite{mycolorado}) 
\item a third party application or website (e.g. Yoti Digital ID \cite{yoti_personal})
\end{itemize}
A Digital ID is a long-term digital representation of an ID.
One key potential benefit of digital IDs, as described in the  ISO/IEC 18013 standard, is their ability to perform  \textit{selective disclosure}, which enables users to pick and choose subset of attributes information to share with relying parties, rather than sharing all ID information \cite{iso-18013-3,iso-18013-5}. The Notary provides an \emph{attestation} of this property. For example, someone could reveal their birthdate, signed by the notary application, to prove their age.
The Assurance on the ID's validity, and the Data Protection of the system, all rely on the Notary: The relying party must trust an attestation about the ID made by the notary, and the User must trust the Notary to handle their ID information privately and securely.
We note that the wallet application where the credential is stored needs not be the same location as where the ID is checked (see \Cref{fig:Verification_location}).
Implementations of this approach include Apple Wallet \cite{applewallet}, Dock Wallet \cite{dock}, Microsoft Entra, \cite{entra}, or a Digital ID application offered by a U.S. State (e.g., \cite{mycolorado} or \cite{NYMID}).

\paragraph{Cryptographic Zero-Knowledge Proofs (ZKPs)} A second family of verifiable ID credentials moves beyond field-level attested disclosure to privacy-preserving cryptography -- removing the need for a trusted notary. When IDs contain a specific authentication tool (specifically, when IDs have digital signatures that can be verified with the \emph{Issuer's} public key) then users can use a cryptographic means to verify (1) that the ID was legitimately issued by the ID Issuer (e.g. the State of New York DMV), (2) the person's attributes were verified (e.g. the ID holder is over 18), and all other information is kept private. Unlike in the Notary case, the sending application does \emph{not} need to be trusted by the verifier -- 
the relying party can use the proof of valid signature from the issuing DMV to check the validity of the ID, and is not relying on the Notary for assurance. 

Implementations of this system rely on interoperable technical standards.
At the data model layer, W3C's Verifiable Credentials data model formalizes a more general-purpose and ``issuer--holder--verifier'' ecosystem supporting cryptographic verifiability \cite{W3C}. Similarly, the SD-JWT VC (IETF-style selective-disclosure JWT credentials) is a practical credential format that preserves JWT compatibility while enabling cryptographically secured statements \cite{SDJWTVC}. At the presentation protocol layer, OpenID for Verifiable Presentations (OpenID4VP) standardizes how wallets present verifiable credentials to services using OAuth/OpenID-style flows \cite{openID}. Further, this approach has been implemented by industrial products and academics, including Google Wallet \cite{googleZKP}, Self \cite{self}, FS'24 \cite{ACECDSA}, RWG+'23 \cite{zk-creds}, ZKPassport \cite{zkpassport}, and LS'25 \cite{conpro25}.

\paragraph{Analysis} Relative to direct ID upload, verifiable ID credentials can materially reduce data exposure by returning only the minimal assertion needed for the transaction, compared to transmitting a full ID. Selective disclosure limits the claims revealed, while ZKPs can further limit disclosure by proving only compliance with a threshold and potentially reducing linkability risks across verifications. Further, verifiable ID credentials support additional features including anti-replay nonces and optional revocation/status checks \citep{W3C}, which further improve security controls and privacy safeguarding. 
However, these approaches introduce new accessibility and operation considerations. First, users may need compatible smartphones, wallets, and connectivity (depending on whether the flow supports offline verification), and recovery processes for lost devices can be complex. Some users (including those without qualifying state IDs, those with limited digital literacy, or those lacking modern devices) may be disadvantaged, suggesting a continued need for fallback mechanisms (e.g., in-person verification or alternative non-ID methods) to avoid exclusion. 
Second, from an implementer perspective, interoperability remains a practical challenge: relying parties must integrate wallet/presentation protocols and align with issuer trust frameworks, which can slow deployment compared to ``upload an ID image'' solutions. Third, usability remains a major challenge. Verifiable digital credentials rely on complex cryptographic mechanisms that can be difficult for non-expert users to understand, which may hinder broader adoption and dilute the practical benefits of their privacy-preserving features.

\subsubsection{Facial age estimation}
Facial age estimation uses models to determine a user’s age by analyzing their face, typically with AI models for regression (age prediction) or classification (classify into age brackets) \cite{elkarazle2022facial}.
The verification flow usually works as follows: first, the user is prompted to enable their camera to capture an image or short video of the user’s face, which could be either processed locally or sent to a cloud API where an AI model performs analysis; second, the model outputs either an age value or a probability that the person is above a certain age \cite{Yotifacial}. The facial age estimation systems rely on large training datasets of facial images labeled with known ages to learn patterns of aging \cite{elkarazle2022facial}.

\paragraph{Analysis.}
Facial age estimation offers a low-friction alternative to ID-based checks and is likely more convenient to consumers than ID-based alternatives. However, they present several material drawbacks that affect both assurance and privacy. First, age inference is inherently difficult, and report shows that age estimation accuracy is ``strongly influenced by algorithm, sex, image quality, region-of-birth, age itself, and interactions between those factors'' \cite{NISTface}. Variations in image quality, makeup, facial expression, and environmental factors such as lighting and camera angle can materially affect model performance \cite{haugg2025imaging}, and potentially result in false rejections of legitimate adults or false acceptances of older-looking minors. Second, accessibility can be a barrier. Users with certain disabilities or medical conditions may struggle with camera-based prompts such as maintaining a steady pose or aligning their face, and individuals with atypical facial features, facial paralysis, or assistive devices may face higher age estimation failure rates~\cite{james2024disabilities}. Finally, even when vendors claim minimal retention, facial age estimation raises privacy concerns because it processes biometric data such as facial images and derived signals that are widely treated as highly sensitive. 

Some facial age estimation deployments additionally include liveness detection, which helps ensure the camera is capturing a real, present user rather than a photo or video replay. Liveness detection can be implemented either by prompting an action (having the user turn their head) or by inferring other signals from the capture (e.g. whether the camera is moving in small ways indicating the camera is held) \cite{liveness}. Action prompting is considered to have stronger assurance, but lower user convenience \cite{liveness}.

\subsection{Where Does the ``Age Checking'' Take Place? On-Device vs.\ Remote}
\label{subsec:verification-location}

\begin{figure*}[h]
    \centering
    \includegraphics[width=\linewidth]{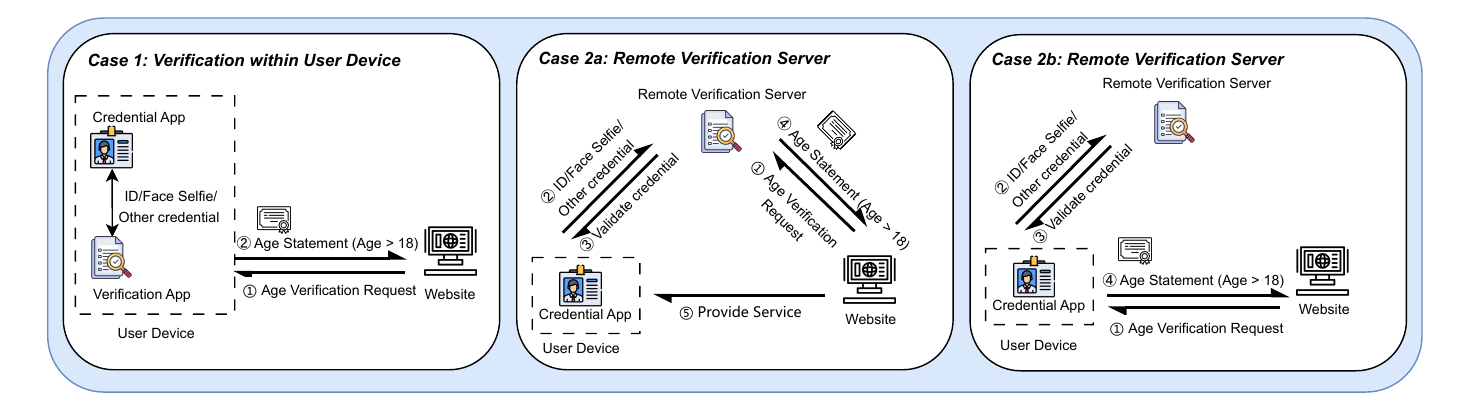}
    \caption{Possibilities for Verification Location: On-Device vs. Remote}
    \label{fig:Verification_location}
\end{figure*}

As we consider these three archetypes, we observe that a key trade-off between \emph{assurance} and \emph{data protection} has to do with 
\emph{where} credential validation and decision logic execute. As illustrated in Figure~\ref{fig:Verification_location}, current systems commonly fall into three deployment scenarios with different privacy, assurance, and operational trade-offs.

\paragraph{Case 1: Verification within the user device.}
In the most privacy-preserving configuration, both the credential storage (wallet/credential app) and the verification logic run locally on the user device, and the website receives only an ``age statement'' (e.g., $\texttt{age>18}$). 
Because raw ID images, selfies, or other credentials need not leave the device, network exposure and third-party retention risks can be reduced. However, on-device checking may constrain assurance: the device may lack access to issuer-side status services (e.g., revocation), may perform weaker authenticity checks than specialized remote infrastructure, and may be more susceptible to tampering on compromised devices. 

\paragraph{Case 2: Remote verification server.}
A common alternative is a remote verification server that receives the credential materials (ID/selfie/other evidence), validates them, and returns an age statement, either directly to the website (case 2a), or back to the user (case 2b), which can then pass on the statement to the website to receive service. 
This configuration can improve assurance by enabling stronger document checks, centralized fraud detection, and consistent policy enforcement by running them on the remote verification server rather than being limited to the capabilities of the user's device. The trade-off is increased risks in data leakage and illegal retention, since sensitive materials traverse the network.

\subsection{Are Credentials Device-Binding or People-Binding?}

A key issue  regarding \emph{assurance} raised in modern age verification systems is the distinction between device-binding and user-binding of credentials. The key distinction lies in what the system ultimately checks: the \emph{device/account} used to access a service, or the \emph{person} physically present at the time of access. We use \emph{device-binding} to denote schemes in which a successful age check results in a persistent authorization state tied primarily to a device identifier, browser token, or account session. Subsequent access is then granted because the same device or account presents the stored token, rather than because the service re-establishes who is currently holding the device. In contrast, \emph{people-binding} denotes schemes in which the age decision is anchored to evidence that the individual present at the time of access satisfies the age requirement, typically via a live biometric check or other person-specific factor that cannot be easily transferred to another user.

\begin{itemize}
    \item \textbf{Direct ID document upload.} Direct ID upload primarily establishes a \emph{document} and \emph{account} binding rather than a robust people-binding, unless an ID and selfie match is required. Still, subsequent access is frequently device- or account-bound because the ``verified'' status is stored as a session token or account attribute. Without re-checks, the approach inherits shared-device vulnerabilities.
    \item \textbf{Verifiable digital ID credentials.} Digital credentials can be either device-bound or people-bound depending on implementation. If a credential is stored in a wallet and released only after local user authentication (PIN/biometric), the flow strengthens people-binding. If credentials are easily re-presented from an unlocked device, or if a long-lived verification token is cached by the relying party, the practical enforcement can become device-bound.
    \item \textbf{Facial age estimation.} Facial age estimation is intended to be people-bound at the point of decision because it evaluates the live user in front of the camera and can incorporate liveness checks to reduce replay.  However, an underage person might simply point the camera at an overage person and cause the ``binding'' to misfire on the wrong person.
\end{itemize}

Device-binding can improve usability by minimizing repeated verification and caching outcomes (e.g., a ``verified for 30 days'' flag), but it is vulnerable to shared-device and shared-account settings. A minor may bypass an age gate by using a parent’s logged-in browser, borrowing an already verified device, or reusing persistent session artifacts. People-binding reduces these ``friendly fraud'' pathways by requiring proof of the eligible individual’s presence at the moment the gated action occurs. However, stronger people-binding can introduce friction (e.g., repeated prompts), create accessibility challenges for some users, and may increase privacy sensitivity if biometrics are involved.

%% file: src/50_Case_study.tex
\section{A Case Study for Analyzing Age Verification Laws: the Tennessee Protecting Children from Social Media Act}
\label{legalanalysis}
With the analytic model discussed in \Cref{model}, and after having discussed technical architectures in \Cref{technical}, we now look into the question of analyzing current age verification laws. 
We emphasize that this discussion does not challenge the reasoning in \textit{FSC v. Paxton} or the general constitutionality of age verification laws.  Rather, it highlights an overlooked but critical concern: even if Texas H.B. 1181 and similar state laws survive constitutional scrutiny, \textit{how severe} will the resulting burdens be? This question is key to answering the question of whether or not the legislation is adequately tailored. That question also depends on the implementation possibilities for age verification and what the frontier of the trade-offs are. This analysis could also be adapted for future laws seeking to verify some non-age information in a similar way.
As a case study, we analyze the Tennessee Protecting Children from Social Media Act.  

\label{sec:case-study}

Tennessee’s Protecting Children from Social Media Act has been effective since January 13, 2025. Similar to other state-level age-verification laws, it establishes an age-gating regime for websites that publish a “substantial portion” of content deemed harmful to minors \cite{Tennessee}. Tennessee also imposes additional operational requirements, including: (1) mandating real-time photo matching between the active user (captured on the access device between the access attempt and viewing) and the photo on a valid state-issued ID \cite{Tennessee}; (2) limiting a verification session to no more than 60 minutes, after which verification must be refreshed \cite{Tennessee}; and (3) requiring verifiers to retain at least 7 years of historical ``anonymized age-verification data,'' which refers to data used to prove the age but ``dissociated with any personally identifying information'' \cite{Tennessee}.

\input{table/Case_Study_Table}

These requirements, particularly photo matching and periodic refresh, more directly advance the Act’s primary target goal of ``protecting children from social media'' as suggested by the Act title. Photo matching introduces an explicit people-binding element that can increase assurance by reducing circumvention via shared devices, borrowed accounts, or reused ID images. The 60-minute session limit further constrains reuse of a single verification event and helps mitigate ongoing access by minors after an adult has verified. Although the fake ID issue discussed in \cref{directID} may still be possible, these design choices still strengthen effective prevention of minors from accessing targeted websites by raising the Assurance level of age gating. However, such achievements come with significant decreases in Data Protection, Consumer Convenience, and Business Convenience, and finally impede Privacy and Security Goal and other Residual Interests. 

First, these photo-matching requirements may produce a substantial chilling effect. A chilling effect arises when ``individuals seeking to engage in activity protected by the First Amendment are deterred from doing so by governmental regulation,'' even when such regulation is not explicitly intended to discourage those activities \cite{schauer1978fear}.
Consumer Convenience is an important evaluative metric in this context. Excessively stringent verification regimes can undermine Consumer Convenience by increasing privacy risks, eroding anonymity, and fostering fear of identity surveillance \cite{user}. Tennessee’s ``photo-matching'' requirements impose multiple layers of burden that are disproportionate to the narrow objective of age verification. Specifically, they require access to highly sensitive biometric data, compel users to entrust government-issued identification to third-party processors, and mandate additional technical steps—such as capturing and uploading live images—that exceed ordinary verification mechanisms. Each incremental requirement introduces additional friction and risk for lawful users, thereby diminishing Consumer Convenience.

Further, the 60-minute verification session further forces users to re-verify their identity repeatedly, disrupting the online experience and creating excessive inconvenience. Rather than enhancing the effectiveness of age-verification regimes, such rules introduce needless friction that can discourage compliance. In sum, when verification regimes demand repeated disclosures of sensitive data, impose short session limits, or require intrusive biometric checks, they substantially increase the burden on ordinary users. These measures reduce Consumer Convenience by introducing friction, privacy risks, and technical obstacles that make lawful access more difficult.

Finally, Tennessee’s requirement to retain seven years of ``ano\-nymized age-verification data'' may materially lower Business Convenience by imposing ongoing compliance, storage, and audit burdens that extend well beyond the real-time act of age gating. First, the definition of ``anonymized age verification data'' and ``personally identifying information'' are unclear or inaccurate. For instance, depending on the technical method used, supposedly ``anonymized'' attributes such as date of birth ranges, device identifiers, or partial government ID numbers may still be re-identifiable when combined with other datasets \cite{rocher2019estimating}, especially over a 7-year window. This definitional ambiguity risks leaving large volumes of quasi-identifiable information vulnerable to breach or misuse. Second, 7 years is a long period for storing any form of sensitive or quasi-sensitive data, especially given the accelerating pace of data breaches and cyberattacks. The longer data are retained, the more likely it is to be exposed through hacking, insider misuse, or even changes in legal regimes that alter the permissible uses of the data. In spite of these privacy concerns, maintaining long-horizon records at scale requires additional logging infrastructure, secure storage and backup, access controls, retention enforcement, and internal governance processes to ensure data remain available, accurate, and defensible over time. These requirements can be especially costly for smaller publishers and for sites that rely on third-party verification vendors, because contractual arrangements, data pipelines, and audit responsibilities must be structured to preserve the mandated records while still complying with minimization obligations. In practice, the combination of long retention and evidentiary expectations increases fixed compliance costs, raises vendor-management complexity, and heightens legal exposure, thereby reducing Business Convenience.

%% file: table/Case_Study_Table.tex
\begin{table*}[ht]
  \centering
  \caption{Case Study of Tennessee: Policy Levers and Their Effects on Variables and Goals}
  \label{tab:policylevers}
  \footnotesize
  \setlength{\tabcolsep}{4pt}
  \resizebox{\linewidth}{!}{%
  \begin{tabular}{@{}p{3.8cm} p{3cm} p{3.5cm} p{3.5cm} p{3.8cm} @{}}
    \toprule
    \textbf{Policy Lever / Design Choice} & \textbf{Relevant Effects} & \textbf{Effect on Target Goal (Protect Minors)} & \textbf{Effect on Privacy \& Security Goal} & \textbf{Effect on Residual Interests (Marketplace Protection, Innovation)} \\
    \midrule
    \textbf{Higher People-binding Thresholds} (e.g., people-binding by ID+ biometric match) & 
    Assurance~\faArrowUp, Business Convenience~\faArrowDown, Consumer Convenience~\faArrowDown, Data Protection~\faArrowDown & 
    Increases effectiveness of age gating $\rightarrow$ Stronger formal alignment with Target Goal & 
    Requires collection of sensitive biometric or ID data; elevates breach and misuse risks & 
    Heightens chilling effects; deters lawful adult access and anonymous speech  \\
    \addlinespace
    \textbf{Short Verification Sessions} (e.g., 60-minute limits) & 
    Assurance~\faArrowUp, Business Convenience~\faArrowDown, Consumer Convenience~\faArrowDown & 
    Mitigating risks of ongoing access by minors $\rightarrow$ Stronger formal alignment with Target Goal & 
    Repeated data submissions amplify exposure risk & 
    Re-verification friction discourages engagement \\
    \addlinespace
    \textbf{Long Data Retention Requirements} (e.g., 7 years) & 
    Business Convenience~\faArrowUp, Consumer Convenience~\faArrowDown, Data Protection~\faArrowDown & 
    N/A & 
    Expands attack surface; conflicts with minimization principles & 
    Raises liability and compliance costs; chills innovation  \\
    \bottomrule
  \end{tabular}
  }
\end{table*}

%% file: src/60_Challenges_Future_Direction.tex
\section{Conclusion}
In sum, age verification in the United States now sits at the intersection of child protection, privacy, free expression, and technological feasibility. The Supreme Court’s decision in \textit{FSC v. Paxton} confirms that age-verification mandates are likely to persist under intermediate scrutiny; therefore laws need to be ``adequately tailored'' to the problem at hand \cite{FreeSpeechCoalition_v_Paxton}. However, the constitutional survival does not end the problem of failing to balance policy tradeoffs among governmental objectives and the need of practical implementation. To solve the remaining problems, this paper proposes a model to interpret  ``adequate tailoring'' among government goals from measurable tradeoffs.
Using the new analysis model, we evaluate current legislation and different verification implementation approaches.
This approach will help with both case-by-case analysis of laws and implementations, and for drawing broader conclusions regarding trade-offs in age verification approaches.

%% file: main.bbl

\begin{thebibliography}{69}


\ifx \showCODEN    \undefined \def \showCODEN     #1{\unskip}     \fi
\ifx \showISBNx    \undefined \def \showISBNx     #1{\unskip}     \fi
\ifx \showISBNxiii \undefined \def \showISBNxiii  #1{\unskip}     \fi
\ifx \showISSN     \undefined \def \showISSN      #1{\unskip}     \fi
\ifx \showLCCN     \undefined \def \showLCCN      #1{\unskip}     \fi
\ifx \shownote     \undefined \def \shownote      #1{#1}          \fi
\ifx \showarticletitle \undefined \def \showarticletitle #1{#1}   \fi
\ifx \showURL      \undefined \def \showURL       {\relax}        \fi
\providecommand\bibfield[2]{#2}
\providecommand\bibinfo[2]{#2}
\providecommand\natexlab[1]{#1}
\providecommand\showeprint[2][]{arXiv:#2}

\bibitem[1092(2024)]%
        {Nebraska}
\bibfield{author}{\bibinfo{person}{Nebraska Legislative~Bill 1092}.} \bibinfo{year}{2024}\natexlab{}.
\newblock
\urldef\tempurl%
\url{https://nebraskalegislature.gov/FloorDocs/108/PDF/Slip/LB1092.pdf}
\showURL{%
\tempurl}


\bibitem[1181(2023)]%
        {Texas}
\bibfield{author}{\bibinfo{person}{Texas House~Bill 1181}.} \bibinfo{year}{2023}\natexlab{}.
\newblock
\urldef\tempurl%
\url{https://capitol.texas.gov/tlodocs/88R/billtext/pdf/HB01181F.pdf#navpanes=0}
\showURL{%
\tempurl}


\bibitem[1515(2023)]%
        {Virginia}
\bibfield{author}{\bibinfo{person}{Virginia Senate~Bill 1515}.} \bibinfo{year}{2023}\natexlab{}.
\newblock
\urldef\tempurl%
\url{https://legacylis.virginia.gov/cgi-bin/legp604.exe?231+ful+CHAP0811}
\showURL{%
\tempurl}


\bibitem[1561(2025)]%
        {Northdakota}
\bibfield{author}{\bibinfo{person}{North Dakota House~Bill 1561}.} \bibinfo{year}{2025}\natexlab{}.
\newblock
\urldef\tempurl%
\url{https://ndlegis.gov/assembly/69-2025/regular/documents/25-0968-03000.pdf}
\showURL{%
\tempurl}


\bibitem[164(2024)]%
        {Alabama}
\bibfield{author}{\bibinfo{person}{Alabama House~Bill 164}.} \bibinfo{year}{2024}\natexlab{}.
\newblock
\urldef\tempurl%
\url{https://alison.legislature.state.al.us/files/pdf/SearchableInstruments/2024RS/HB164-enr.pdf}
\showURL{%
\tempurl}


\bibitem[17(2024)]%
        {Indiana}
\bibfield{author}{\bibinfo{person}{Indiana Senate~Bill 17}.} \bibinfo{year}{2024}\natexlab{}.
\newblock
\urldef\tempurl%
\url{https://iga.in.gov/pdf-documents/123/2024/senate/bills/SB0017/SB0017.05.ENRH.pdf}
\showURL{%
\tempurl}


\bibitem[1792(2024)]%
        {Tennessee}
\bibfield{author}{\bibinfo{person}{Tennessee Senate~Bill 1792}.} \bibinfo{year}{2024}\natexlab{}.
\newblock
\urldef\tempurl%
\url{https://publications.tnsosfiles.com/acts/113/pub/pc1021.pdf}
\showURL{%
\tempurl}


\bibitem[1959(2024)]%
        {Oklahoma}
\bibfield{author}{\bibinfo{person}{Oklahoma Senate~Bill 1959}.} \bibinfo{year}{2024}\natexlab{}.
\newblock
\urldef\tempurl%
\url{https://www.oklegislature.gov/cf_pdf/2023-24%20ENR/SB/SB1959%20ENR.PDF}
\showURL{%
\tempurl}


\bibitem[2112(2025)]%
        {Arizona}
\bibfield{author}{\bibinfo{person}{Arizona House~Bill 2112}.} \bibinfo{year}{2025}\natexlab{}.
\newblock
\urldef\tempurl%
\url{https://www.azleg.gov/legtext/57leg/1R/laws/0193.pdf}
\showURL{%
\tempurl}


\bibitem[2346(2024)]%
        {Mississippi}
\bibfield{author}{\bibinfo{person}{Senate~Bill 2346}.} \bibinfo{year}{2024}\natexlab{}.
\newblock
\urldef\tempurl%
\url{https://billstatus.ls.state.ms.us/2023/pdf/history/SB/SB2346.xml}
\showURL{%
\tempurl}


\bibitem[278(2024)]%
        {Kentucky}
\bibfield{author}{\bibinfo{person}{Kentucky House~Bill 278}.} \bibinfo{year}{2024}\natexlab{}.
\newblock
\urldef\tempurl%
\url{https://apps.legislature.ky.gov/law/acts/24RS/documents/0106.pdf}
\showURL{%
\tempurl}


\bibitem[3(2024)]%
        {Florida}
\bibfield{author}{\bibinfo{person}{Florida House~Bill 3}.} \bibinfo{year}{2024}\natexlab{}.
\newblock
\urldef\tempurl%
\url{https://www.flsenate.gov/Session/Bill/2024/3/BillText/er/PDF}
\showURL{%
\tempurl}


\bibitem[3424(2024a)]%
        {Southcarolina}
\bibfield{author}{\bibinfo{person}{South Carolina~H 3424}.} \bibinfo{year}{2024}\natexlab{a}.
\newblock
\urldef\tempurl%
\url{https://www.scstatehouse.gov/sess125_2023-2024/bills/3424.htm}
\showURL{%
\tempurl}


\bibitem[3424(2024b)]%
        {Southdakota}
\bibfield{author}{\bibinfo{person}{South Dakota House~Bill 3424}.} \bibinfo{year}{2024}\natexlab{b}.
\newblock
\urldef\tempurl%
\url{https://sdlegislature.gov/Session/Bill/25525/282508}
\showURL{%
\tempurl}


\bibitem[351(2024)]%
        {Georgia}
\bibfield{author}{\bibinfo{person}{Georgia Senate~Bill 351}.} \bibinfo{year}{2024}\natexlab{}.
\newblock
\urldef\tempurl%
\url{https://www.legis.ga.gov/legislation/66023}
\showURL{%
\tempurl}


\bibitem[394(2024)]%
        {Kansas}
\bibfield{author}{\bibinfo{person}{Kansas Senate Bill~No. 394}.} \bibinfo{year}{2024}\natexlab{}.
\newblock
\urldef\tempurl%
\url{https://www.kslegislature.gov/li_2024/b2023_24/measures/documents/sb394_enrolled.pdf}
\showURL{%
\tempurl}


\bibitem[43(2025)]%
        {Wyoming}
\bibfield{author}{\bibinfo{person}{Wyoming House~Bill 43}.} \bibinfo{year}{2025}\natexlab{}.
\newblock
\urldef\tempurl%
\url{https://wyoleg.gov/Legislation/2025/HB0043}
\showURL{%
\tempurl}


\bibitem[498(2024)]%
        {Idaho}
\bibfield{author}{\bibinfo{person}{Idaho House~Bill 498}.} \bibinfo{year}{2024}\natexlab{}.
\newblock
\urldef\tempurl%
\url{https://legislature.idaho.gov/wp-content/uploads/sessioninfo/2024/legislation/H0498.pdf}
\showURL{%
\tempurl}


\bibitem[544(2023)]%
        {Montana}
\bibfield{author}{\bibinfo{person}{Montana Senate~Bill 544}.} \bibinfo{year}{2023}\natexlab{}.
\newblock
\urldef\tempurl%
\url{https://s3.amazonaws.com/fn-document-service/file-by-sha384/b618d2b4851fe7516875fd9380bea85a9ab91f2a21a130b49fed787a6163b1139310b43582b30c2ff68d058625ee48e4}
\showURL{%
\tempurl}


\bibitem[60-18(2025)]%
        {Missouri}
\bibfield{author}{\bibinfo{person}{Missouri 15~CSR 60-18}.} \bibinfo{year}{2025}\natexlab{}.
\newblock
\urldef\tempurl%
\url{https://assets.freespeechcoalition.com/documents/MO%20AV%20Rule.pdf}
\showURL{%
\tempurl}


\bibitem[66(2023)]%
        {Arkansas}
\bibfield{author}{\bibinfo{person}{Arkansas Senate~Bill 66}.} \bibinfo{year}{2023}\natexlab{}.
\newblock
\urldef\tempurl%
\url{https://arkleg.state.ar.us/Home/FTPDocument?path=%2FACTS%2F2023R%2FPublic%2FACT612.pdf}
\showURL{%
\tempurl}


\bibitem[8(2024)]%
        {NorthCarolina}
\bibfield{author}{\bibinfo{person}{North Carolina House~Bill 8}.} \bibinfo{year}{2024}\natexlab{}.
\newblock
\urldef\tempurl%
\url{https://www.ncleg.gov/Sessions/2023/Bills/House/PDF/H8v5.pdf}
\showURL{%
\tempurl}


\bibitem[96(2025)]%
        {Ohio}
\bibfield{author}{\bibinfo{person}{Ohio House~Bill 96}.} \bibinfo{year}{2025}\natexlab{}.
\newblock
\urldef\tempurl%
\url{https://codes.ohio.gov/ohio-revised-code/section-1349.10}
\showURL{%
\tempurl}


\bibitem[Alvargonzález(2025)]%
        {liveness}
\bibfield{author}{\bibinfo{person}{Andrés Alvargonzález}.} \bibinfo{year}{2025}\natexlab{}.
\newblock \bibinfo{title}{What is a liveness check and why it matters in facial biometrics}.
\newblock \bibinfo{howpublished}{Web page}.
\newblock
\urldef\tempurl%
\url{https://www.identy.io/what-is-a-liveness-check-and-why-it-matters-in-facial-biometrics/}
\showURL{%
\tempurl}
\newblock
\shownote{Accessed: 2026-01-09}.


\bibitem[Anduze(2024)]%
        {anduze2024obscenity}
\bibfield{author}{\bibinfo{person}{Daniel~S Anduze}.} \bibinfo{year}{2024}\natexlab{}.
\newblock \showarticletitle{Obscenity Revisited: Defending Recent Age-Verification Laws Against First Amendment Challenges}.
\newblock \bibinfo{journal}{\emph{Colum. JL \& Soc. Probs.}}  \bibinfo{volume}{58} (\bibinfo{year}{2024}), \bibinfo{pages}{147}.
\newblock


\bibitem[{Apple Inc.}(2025)]%
        {applewallet}
\bibfield{author}{\bibinfo{person}{{Apple Inc.}}} \bibinfo{year}{2025}\natexlab{}.
\newblock \bibinfo{title}{Add your driver's license to Apple Wallet}.
\newblock
\newblock
\shownote{\url{https://support.apple.com/en-us/111803}}.


\bibitem[Boult(2025)]%
        {detectfakeID}
\bibfield{author}{\bibinfo{person}{Luke~Owain Boult}.} \bibinfo{year}{2025}\natexlab{}.
\newblock \bibinfo{title}{How to Spot a Fake ID in the US by State: Red Flags \& Verification Tools}.
\newblock \bibinfo{howpublished}{Web page}.
\newblock
\urldef\tempurl%
\url{https://sumsub.com/blog/how-to-spot-fake-id-us/}
\showURL{%
\tempurl}
\newblock
\shownote{Accessed: 2026-01-09}.


\bibitem[Center(2025a)]%
        {Freespeech}
\bibfield{author}{\bibinfo{person}{FSC~Action Center}.} \bibinfo{year}{2025}\natexlab{a}.
\newblock \bibinfo{title}{Age Verification: The Complicated Effort to Protect Youth Online}.
\newblock
\newblock
\shownote{\url{https://action.freespeechcoalition.com/age-verification-resources/state-avs-laws/}}.


\bibitem[Center(2025b)]%
        {FSClitigation}
\bibfield{author}{\bibinfo{person}{FSC~Action Center}.} \bibinfo{year}{2025}\natexlab{b}.
\newblock \bibinfo{title}{Cases in Progress}.
\newblock
\newblock
\shownote{\url{https://action.freespeechcoalition.com/age-verification-resources/av-lawsuits/}}.


\bibitem[Cross(2004)]%
        {cross2004age}
\bibfield{author}{\bibinfo{person}{John~T Cross}.} \bibinfo{year}{2004}\natexlab{}.
\newblock \showarticletitle{Age verification in the 21st century: Swiping away your privacy}.
\newblock \bibinfo{journal}{\emph{J. Marshall J. Computer \& Info. L.}}  \bibinfo{volume}{23} (\bibinfo{year}{2004}), \bibinfo{pages}{363}.
\newblock


\bibitem[Dahlstrom(2023)]%
        {dahlstrom2023new}
\bibfield{author}{\bibinfo{person}{Julie Dahlstrom}.} \bibinfo{year}{2023}\natexlab{}.
\newblock \showarticletitle{The new pornography wars}.
\newblock \bibinfo{journal}{\emph{Fla. L. Rev.}}  \bibinfo{volume}{75} (\bibinfo{year}{2023}), \bibinfo{pages}{117}.
\newblock


\bibitem[{Dock Labs}(2024)]%
        {dock}
\bibfield{author}{\bibinfo{person}{{Dock Labs}}.} \bibinfo{year}{2024}\natexlab{}.
\newblock \bibinfo{title}{Selective Disclosure: Choose What Data To Share}.
\newblock
\newblock
\shownote{\url{https://www.dock.io/post/selective-disclosure}}.


\bibitem[ELKarazle et~al\mbox{.}(2022)]%
        {elkarazle2022facial}
\bibfield{author}{\bibinfo{person}{Khaled ELKarazle}, \bibinfo{person}{Valliappan Raman}, {and} \bibinfo{person}{Patrick Then}.} \bibinfo{year}{2022}\natexlab{}.
\newblock \showarticletitle{Facial age estimation using machine learning techniques: An overview}.
\newblock \bibinfo{journal}{\emph{Big Data and Cognitive Computing}} \bibinfo{volume}{6}, \bibinfo{number}{4} (\bibinfo{year}{2022}), \bibinfo{pages}{128}.
\newblock


\bibitem[Fisher and Galluzzo(2024)]%
        {verifiabledigitalcredential}
\bibfield{author}{\bibinfo{person}{Bill Fisher} {and} \bibinfo{person}{Ryan Galluzzo}.} \bibinfo{year}{2024}\natexlab{}.
\newblock \bibinfo{title}{Digital Identities: Getting to Know the Verifiable Digital Credential Ecosystem}.
\newblock \bibinfo{howpublished}{Web page}.
\newblock
\urldef\tempurl%
\url{https://www.nist.gov/blogs/cybersecurity-insights/digital-identities-getting-know-verifiable-digital-credential-ecosystem}
\showURL{%
\tempurl}
\newblock
\shownote{Accessed: 2026-01-09}.


\bibitem[Frigo and abhi shelat(2024)]%
        {ACECDSA}
\bibfield{author}{\bibinfo{person}{Matteo Frigo} {and} \bibinfo{person}{abhi shelat}.} \bibinfo{year}{2024}\natexlab{}.
\newblock \bibinfo{title}{Anonymous credentials from {ECDSA}}.
\newblock \bibinfo{howpublished}{Cryptology {ePrint} Archive, Paper 2024/2010}.
\newblock
\urldef\tempurl%
\url{https://eprint.iacr.org/2024/2010}
\showURL{%
\tempurl}


\bibitem[Frischmann et~al\mbox{.}(2024)]%
        {frischmann2024online}
\bibfield{author}{\bibinfo{person}{Brett~M Frischmann}, \bibinfo{person}{Noah Apthorpe}, {and} \bibinfo{person}{Yan Shvartzshnaider}.} \bibinfo{year}{2024}\natexlab{}.
\newblock \bibinfo{title}{Online Age Gating: An Interdisciplinary Evaluation}.
\newblock
\newblock
\shownote{Available at SSRN 4937328}.


\bibitem[{Google}(2024)]%
        {googleZKP}
\bibfield{author}{\bibinfo{person}{{Google}}.} \bibinfo{year}{2024}\natexlab{}.
\newblock \bibinfo{booktitle}{\emph{New ways to verify your age and identity with {Google Wallet}}}.
\newblock Google.
\newblock
\urldef\tempurl%
\url{https://blog.google/products/google-pay/google-wallet-age-identity-verifications/}
\showURL{%
\tempurl}
\newblock
\shownote{Accessed: 2025-07-14}.


\bibitem[Hanaoka et~al\mbox{.}(2024)]%
        {NISTface}
\bibfield{author}{\bibinfo{person}{Kayee Hanaoka}, \bibinfo{person}{Mei Ngan}, \bibinfo{person}{Joyce Yang}, \bibinfo{person}{George~W. Quinn}, \bibinfo{person}{Austin Hom}, {and} \bibinfo{person}{Patrick Grother}.} \bibinfo{year}{2024}\natexlab{}.
\newblock \bibinfo{booktitle}{\emph{Face Analysis Technology Evaluation: Age Estimation and Verification}}.
\newblock \bibinfo{type}{NIST Interagency Report} 8525. \bibinfo{institution}{National Institute of Standards and Technology}, \bibinfo{address}{Gaithersburg, MD}.
\newblock
\href{https://doi.org/10.6028/NIST.IR.8525}{doi:\nolinkurl{10.6028/NIST.IR.8525}}
\newblock
\shownote{Approved by the NIST Editorial Review Board on 2024-05-24; report includes updates as of 2025-12-18.}.


\bibitem[Haugg et~al\mbox{.}(2025)]%
        {haugg2025imaging}
\bibfield{author}{\bibinfo{person}{Fridolin Haugg}, \bibinfo{person}{Grace Lee}, \bibinfo{person}{John He}, \bibinfo{person}{Justin Johnson}, \bibinfo{person}{Anna Zapaishchykova}, \bibinfo{person}{Danielle~S Bitterman}, \bibinfo{person}{Benjamin~H Kann}, \bibinfo{person}{Hugo~JWL Aerts}, {and} \bibinfo{person}{Raymond~H Mak}.} \bibinfo{year}{2025}\natexlab{}.
\newblock \showarticletitle{Imaging biomarkers of ageing: a review of artificial intelligence-based approaches for age estimation}.
\newblock \bibinfo{journal}{\emph{The Lancet Healthy Longevity}} (\bibinfo{year}{2025}).
\newblock


\bibitem[{International Organization for Standardization}(2017)]%
        {iso-18013-3}
\bibfield{author}{\bibinfo{person}{{International Organization for Standardization}}.} \bibinfo{year}{2017}\natexlab{}.
\newblock \bibinfo{title}{Personal identification — ISO-compliant driving licence Part 3: Access control, authentication and integrity validation}.
\newblock
\newblock
\shownote{https://www.iso.org/standard/72366.html}.


\bibitem[{International Organization for Standardization}(2021)]%
        {iso-18013-5}
\bibfield{author}{\bibinfo{person}{{International Organization for Standardization}}.} \bibinfo{year}{2021}\natexlab{}.
\newblock \bibinfo{title}{Personal identification — ISO-compliant driving licence Part 5: Mobile driving licence (mDL) application}.
\newblock
\newblock
\shownote{https://www.iso.org/standard/69084.html}.


\bibitem[James(2024)]%
        {james2024disabilities}
\bibfield{author}{\bibinfo{person}{Ness James}.} \bibinfo{year}{2024}\natexlab{}.
\newblock \bibinfo{booktitle}{\emph{Access to services: The promises and pitfalls of AI for people with disabilities}}.
\newblock European Disability Forum.
\newblock
\urldef\tempurl%
\url{https://www.edf-feph.org/access-to-services-the-promises-and-pitfalls-of-ai-for-people-with-disabilities/}
\showURL{%
\tempurl}


\bibitem[La. Acts No. 440, § 1(2023)]%
        {Louisiana}
La. Acts No. 440, § 1 \bibinfo{year}{2023}\natexlab{}.
\newblock
\urldef\tempurl%
\url{https://www.legis.la.gov/Legis/ViewDocument.aspx?d=1289498}
\showURL{%
\tempurl}


\bibitem[Ligon(2025)]%
        {idscan_ocr}
\bibfield{author}{\bibinfo{person}{Hannah Ligon}.} \bibinfo{year}{2025}\natexlab{}.
\newblock \bibinfo{title}{What is {OCR} in {ID} and Document Scanning?}
\newblock \bibinfo{howpublished}{Web page}.
\newblock
\urldef\tempurl%
\url{https://idscan.net/blog/what-is-ocr-in-id-and-document-scanning/}
\showURL{%
\tempurl}
\newblock
\shownote{Accessed: 2026-01-09}.


\bibitem[Lin et~al\mbox{.}(2025)]%
        {user}
\bibfield{author}{\bibinfo{person}{Yanzi~Veronica Lin}, \bibinfo{person}{Vivianna Lieu}, \bibinfo{person}{Cheng Zhang}, \bibinfo{person}{Weiqian Zhang}, \bibinfo{person}{Wenchao Hu}, \bibinfo{person}{Lorrie~Faith Cranor}, {and} \bibinfo{person}{Sarah Scheffler}.} \bibinfo{year}{2025}\natexlab{}.
\newblock \showarticletitle{Carded by the Internet: Measuring User Responses to Online Age Assurance Mechanisms}.
\newblock \bibinfo{journal}{\emph{USENIX SOUPS Poster}} (\bibinfo{year}{2025}).
\newblock


\bibitem[Liu and Scheffler(2025)]%
        {conpro25}
\bibfield{author}{\bibinfo{person}{Shuang Liu} {and} \bibinfo{person}{Sarah Scheffler}.} \bibinfo{year}{2025}\natexlab{}.
\newblock \bibinfo{title}{Privacy-preserving Age Verification based on Improved Verifiable Credentials Framework}.
\newblock
\newblock
\shownote{https://conpro25.ieee-security.org/papers/liu-conpro25.pdf}.


\bibitem[Marsden(2023)]%
        {marsden2023age}
\bibfield{author}{\bibinfo{person}{Christine Marsden}.} \bibinfo{year}{2023}\natexlab{}.
\newblock \showarticletitle{Age-Verification Laws in the Era of Digital Privacy}.
\newblock \bibinfo{journal}{\emph{Nat'l Sec. LJ}}  \bibinfo{volume}{10} (\bibinfo{year}{2023}), \bibinfo{pages}{210}.
\newblock


\bibitem[MATTR et~al\mbox{.}(2025)]%
        {openID}
\bibfield{author}{\bibinfo{person}{O.~Terbu MATTR}, \bibinfo{person}{T.~Lodderstedt SPRIND}, \bibinfo{person}{K.~Yasuda SPRIND}, \bibinfo{person}{D.~Fett Authlete}, {and} \bibinfo{person}{J.~Heenan Authlete}.} \bibinfo{year}{2025}\natexlab{}.
\newblock \bibinfo{title}{OpenID for Verifiable Presentations 1.0}.
\newblock \bibinfo{howpublished}{Web page}.
\newblock
\urldef\tempurl%
\url{https://openid.net/specs/openid-4-verifiable-presentations-1_0.html}
\showURL{%
\tempurl}
\newblock
\shownote{Accessed: 2026-01-09}.


\bibitem[{Microsoft}(2025)]%
        {entra}
\bibfield{author}{\bibinfo{person}{{Microsoft}}.} \bibinfo{year}{2025}\natexlab{}.
\newblock \bibinfo{booktitle}{\emph{Introduction to Microsoft Entra Verified ID}}.
\newblock
\urldef\tempurl%
\url{https://learn.microsoft.com/en-us/entra/verified-id/decentralized-identifier-overview}
\showURL{%
\tempurl}
\newblock
\shownote{Accessed on 2025-08-07}.


\bibitem[Mithani(2025)]%
        {DFnews3}
\bibfield{author}{\bibinfo{person}{Jasmine Mithani}.} \bibinfo{year}{2025}\natexlab{}.
\newblock \bibinfo{booktitle}{\emph{Kids are making deepfakes of each other, and laws aren’t keeping up}}.
\newblock
\urldef\tempurl%
\url{https://19thnews.org/2025/07/deepfake-ai-kids-schools-laws-policy/}
\showURL{%
Retrieved September 25, 2025 from \tempurl}


\bibitem[of~Motor Vehicle~Administrators(2025)]%
        {AAMVAstandards}
\bibfield{author}{\bibinfo{person}{American~Association of Motor Vehicle~Administrators}.} \bibinfo{year}{2025}\natexlab{}.
\newblock \bibinfo{title}{AAMVA DL/ID Card Design Standard}.
\newblock \bibinfo{howpublished}{Web page}.
\newblock
\urldef\tempurl%
\url{https://www.aamva.org/getmedia/81af105d-8b1b-45e1-aa46-f1800a259ed1/AAMVADLIDCardDesignStandard2025.pdf}
\showURL{%
\tempurl}
\newblock
\shownote{Accessed: 2026-01-09}.


\bibitem[of~Motor~Vehicles({[n.\,d.]})]%
        {NYMID}
\bibfield{author}{\bibinfo{person}{New York~Department of Motor~Vehicles}.} \bibinfo{year}{[n.\,d.]}\natexlab{}.
\newblock \bibinfo{booktitle}{\emph{Mobile ID (MiD)}}.
\newblock
\urldef\tempurl%
\url{https://dmv.ny.gov/id-card/mobile-id-mid}
\showURL{%
Retrieved March 03, 2025 from \tempurl}


\bibitem[Oliver~Terbu(2025)]%
        {SDJWTVC}
\bibfield{author}{\bibinfo{person}{Brian~Campbell Oliver~Terbu, Daniel~Fett}.} \bibinfo{year}{2025}\natexlab{}.
\newblock \bibinfo{title}{SD-JWT-based Verifiable Credentials (SD-JWT VC)}.
\newblock \bibinfo{howpublished}{Web page}.
\newblock
\urldef\tempurl%
\url{https://datatracker.ietf.org/doc/draft-ietf-oauth-sd-jwt-vc/}
\showURL{%
\tempurl}
\newblock
\shownote{Accessed: 2026-01-09}.


\bibitem[Reno v. Am. C.L. Union, 521 U.S. 844, 882(1997)]%
        {Reno}
Reno v. Am. C.L. Union, 521 U.S. 844, 882 \bibinfo{year}{1997}\natexlab{}.
\newblock


\bibitem[Rocher et~al\mbox{.}(2019)]%
        {rocher2019estimating}
\bibfield{author}{\bibinfo{person}{Luc Rocher}, \bibinfo{person}{Julien~M Hendrickx}, {and} \bibinfo{person}{Yves-Alexandre De~Montjoye}.} \bibinfo{year}{2019}\natexlab{}.
\newblock \showarticletitle{Estimating the success of re-identifications in incomplete datasets using generative models}.
\newblock \bibinfo{journal}{\emph{Nature communications}} \bibinfo{volume}{10}, \bibinfo{number}{1} (\bibinfo{year}{2019}), \bibinfo{pages}{3069}.
\newblock


\bibitem[Rosenberg et~al\mbox{.}(2023)]%
        {zk-creds}
\bibfield{author}{\bibinfo{person}{Michael Rosenberg}, \bibinfo{person}{Jacob White}, \bibinfo{person}{Christina Garman}, {and} \bibinfo{person}{Ian Miers}.} \bibinfo{year}{2023}\natexlab{}.
\newblock \showarticletitle{zk-creds: Flexible anonymous credentials from zksnarks and existing identity infrastructure}. In \bibinfo{booktitle}{\emph{2023 IEEE Symposium on Security and Privacy (SP)}}. IEEE, \bibinfo{pages}{790--808}.
\newblock


\bibitem[S.B.287(2023)]%
        {Utah}
\bibfield{author}{\bibinfo{person}{Utah S.B.287}.} \bibinfo{year}{2023}\natexlab{}.
\newblock
\urldef\tempurl%
\url{https://le.utah.gov/~2023/bills/static/SB0287.html}
\showURL{%
\tempurl}


\bibitem[Schauer(1978)]%
        {schauer1978fear}
\bibfield{author}{\bibinfo{person}{Frederick Schauer}.} \bibinfo{year}{1978}\natexlab{}.
\newblock \showarticletitle{{Fear, Risk and the First Amendment: Unraveling the "Chilling Effect"}}.
\newblock \bibinfo{journal}{\emph{{Boston University Law Review}}}  \bibinfo{volume}{58} (\bibinfo{year}{1978}), \bibinfo{pages}{685--732}.
\newblock


\bibitem[{Self Docs}(2025)]%
        {self}
\bibfield{author}{\bibinfo{person}{{Self Docs}}.} \bibinfo{year}{2025}\natexlab{}.
\newblock \bibinfo{booktitle}{\emph{Self Protocol}}.
\newblock
\urldef\tempurl%
\url{https://docs.self.xyz/}
\showURL{%
\tempurl}
\newblock
\shownote{Last updated July 21, 2025}.


\bibitem[Singer(2024)]%
        {NYTDFnews}
\bibfield{author}{\bibinfo{person}{Natasha Singer}.} \bibinfo{year}{2024}\natexlab{}.
\newblock \bibinfo{booktitle}{\emph{Teen Girls Confront an Epidemic of Deepfake Nudes in Schools}}.
\newblock
\urldef\tempurl%
\url{https://www.nytimes.com/2024/04/08/technology/deepfake-ai-nudes-westfield-high-school.html}
\showURL{%
Retrieved September 25, 2025 from \tempurl}


\bibitem[Sporny et~al\mbox{.}(2025)]%
        {W3C}
\bibfield{author}{\bibinfo{person}{Manu Sporny}, \bibinfo{person}{Dave Longley}, \bibinfo{person}{David Chadwick}, {and} \bibinfo{person}{Ivan Herman}.} \bibinfo{year}{2025}\natexlab{}.
\newblock \bibinfo{booktitle}{\emph{Verifiable Credentials Data Model v2.0}}.
\newblock
\urldef\tempurl%
\url{https://www.w3.org/TR/vc-data-model-2.0/#dfn-verifiable-credential}
\showURL{%
Retrieved February 22, 2025 from \tempurl}


\bibitem[{State of Colorado}(2025)]%
        {mycolorado}
\bibfield{author}{\bibinfo{person}{{State of Colorado}}.} \bibinfo{year}{2025}\natexlab{}.
\newblock \bibinfo{title}{{myColorado} State of Colorado's Official Mobile App}.
\newblock
\newblock
\shownote{\url{https://mycolorado.gov/colorado-digital-id/verify}}.


\bibitem[{Supreme Court of the United States}(2025)]%
        {FreeSpeechCoalition_v_Paxton}
\bibfield{author}{\bibinfo{person}{{Supreme Court of the United States}}.} \bibinfo{year}{2025}\natexlab{}.
\newblock \bibinfo{title}{Free Speech Coalition, Inc. v. Paxton}.
\newblock \bibinfo{howpublished}{U.S. Supreme Court, Docket No. 23-1122}.
\newblock
\urldef\tempurl%
\url{https://www.supremecourt.gov/opinions/24pdf/23-1122_3e04.pdf}
\showURL{%
\tempurl}
\newblock
\shownote{Argued Jan. 15, 2025; Decided Jun. 27, 2025. Appeal from the Fifth Circuit.}.


\bibitem[Thierer(2007)]%
        {Thierer2004}
\bibfield{author}{\bibinfo{person}{Adam Thierer}.} \bibinfo{year}{2007}\natexlab{}.
\newblock \showarticletitle{Social Networking and Age Verification: Many Hard Questions; No Easy Solutions}.
\newblock \bibinfo{journal}{\emph{SSRN Electronic Journal}} (\bibinfo{date}{03} \bibinfo{year}{2007}).
\newblock
\href{https://doi.org/10.2139/ssrn.976936}{doi:\nolinkurl{10.2139/ssrn.976936}}


\bibitem[Winnard(2024)]%
        {DFnews2}
\bibfield{author}{\bibinfo{person}{Nigel Winnard}.} \bibinfo{year}{2024}\natexlab{}.
\newblock \bibinfo{booktitle}{\emph{The Rise of Deepfakes in Schools}}.
\newblock
\urldef\tempurl%
\url{https://www.tieonline.com/article/3632/the-rise-of-deepfakes-in-schools}
\showURL{%
Retrieved September 25, 2025 from \tempurl}


\bibitem[Yoti(2025a)]%
        {Yotifacial}
\bibfield{author}{\bibinfo{person}{Yoti}.} \bibinfo{year}{2025}\natexlab{a}.
\newblock \bibinfo{title}{Facial age estimation}.
\newblock \bibinfo{howpublished}{Web page}.
\newblock
\urldef\tempurl%
\url{https://www.yoti.com/business/facial-age-estimation/}
\showURL{%
\tempurl}
\newblock
\shownote{Accessed: 2026-01-09}.


\bibitem[Yoti(2025b)]%
        {yoti_personal}
\bibfield{author}{\bibinfo{person}{Yoti}.} \bibinfo{year}{2025}\natexlab{b}.
\newblock \bibinfo{title}{Yoti ID is your secure Digital ID}.
\newblock \bibinfo{howpublished}{\url{https://www.yoti.com/personal/}}.
\newblock
\newblock
\shownote{Accessed: 2026-01-13. Yoti is a secure digital identity service that lets individuals store their personal details, prove identity and age, and securely share selected identity information with others.}.


\bibitem[Zipursky(2015)]%
        {zipursky2015reasonableness}
\bibfield{author}{\bibinfo{person}{Benjamin~C Zipursky}.} \bibinfo{year}{2015}\natexlab{}.
\newblock \showarticletitle{Reasonableness In and Out of Negligence Law}.
\newblock \bibinfo{journal}{\emph{University of Pennsylvania Law Review}}  \bibinfo{volume}{163} (\bibinfo{year}{2015}), \bibinfo{pages}{2131}.
\newblock


\bibitem[{ZKPassport}(2025)]%
        {zkpassport}
\bibfield{author}{\bibinfo{person}{{ZKPassport}}.} \bibinfo{year}{2025}\natexlab{}.
\newblock \bibinfo{booktitle}{\emph{{ZKPassport} Introduction}}.
\newblock
\urldef\tempurl%
\url{https://docs.zkpassport.id/intro}
\showURL{%
\tempurl}
\newblock
\shownote{Accessed: 2025-08-07}.


\end{thebibliography}
